\documentclass[prd,aps,twocolumn]{revtex4}
\usepackage{graphicx}

\def\gr{$\gamma$-ray}

\begin{document}
\title{LHAASO sensitivity for diffuse gamma-ray signals from the Galaxy.}
\author{Andrii Neronov$^{1,2}$, Dmitri Semikoz$^1$}
\affiliation{$^1$Université de Paris,
CNRS, Astroparticule et Cosmologie, F-75013 Paris, France\\
$^2$Astronomy Department, University of Geneva, Ch. d'Ecogia 16, 1290, Versoix, Switzerland}

\begin{abstract}
 We estimate the sensitivity of LHAASO telescope for the large angular scale diffuse \gr\ flux in multi-TeV -- multi-PeV energy range. We discuss possible sources of the signal in this energy range including the guaranteed flux from cosmic ray interactions in the interstellar medium and possible flux from decaying dark matter. We show that LHAASO will be able to detect the diffuse cosmic ray induced \gr\ flux up to high Galactic latitude regions thus providing firm identification of the Galactic cosmic ray component of the astrophysical neutrino signal detected by IceCube and clarification of the nature of the knee feature of the cosmic ray spectrum. Comparing the diffuse flux sensitivity with the diffuse \gr\ flux expected from the dark matter decays, we show LHAASO will be able to detect the \gr\ signal from dark matter particles of PeV-EeV mass decaying on the time scale up to $3\times 10^{29}$~s. 
\end{abstract}
\maketitle

%%%%%%%%%%%%%%%%%%%%%%%%%%%%%%%%%%%%%%%%%%
\section{Introduction}
%%%%%%%%%%%%%%%%%%%%%%%%%%%%%%%%%%%%%%%%%%

Diffuse \gr\ flux from cosmic ray interactions in the Milky Way galaxy provides the strongest \gr\ signal on the sky \cite{fermi_diffuse_2012,neronov_semikoz19,hess_diffuse}. This signal is  measured up to the energy $\sim 3$~TeV all across the Galactic Plane, in the mid and high Galactic latitude regions by Fermi/LAT telescopes \cite{atwood09,fermi_diffuse_2012,neronov_semikoz19}. Its spectrum at the highest energies is consistent with a powerlaw $dN_\gamma/dE\propto E^{-\Gamma_\gamma}$ with the slope $\Gamma_\gamma\simeq 2.4$, with no signature of high-energy cut-off. 

The main source of Galactic \gr s in the multi-TeV energy band is  decays of pions resulting from interactions of cosmic rays in the interstellar medium. Other diffuse flux components, such as the extragalactic \gr\ flux and inverse Compton emission from cosmic ray electrons are suppressed in this  energy range. The extragalactic photons could not reach the Earth because of the pair production on the Extragalactic Background Light (EBL) \cite{gould,franceschini08}. The inverse Compton flux is suppressed because of suppression of the scattering cross-section in the Klein-Nishina regime of Compton scattering of the interstellar radiation field photons \cite{blumenthal,schlickeiser} and due to the high-energy cut-off in the spectrum of cosmic ray electrons  \cite{hess_icrc2017}.

The pion decay photons carry on average a fraction $E_\gamma\sim \kappa E_p,\  \kappa\simeq 0.04\ll 1$ of the parent proton energy $E_p$ \cite{kelner06,kappes}. This suggests that the Galactic diffuse emission spectrum is expected to ultimately have a high-energy softening at the energy by a factor $\kappa$ lower of the limiting energy at which cosmic ray protons (and atomic nuclei) could not anymore be retained by the Galactic magnetic field \cite{smirnov}. If this characteristic energy is in the range of the "knee" of the cosmic ray spectrum at 1-10 PeV, the spectrum of the cosmic-ray generated diffuse \gr\ flux is expected to have a soften in the energy range 10-100 TeV. Measurement of such softening at different locations across the Galaxy is, in principle, possible with \gr\ telescopes sensitive in the TeV-PeV energy range. Such measurement would provide an important step toward understanding of the mechanism of propagation of cosmic rays through the interstellar medium and escape of cosmic rays from the Milky Way. 

Apart from the conventional cosmic ray induced \gr\ flux, the TeV-PeV band diffuse emission might contain new types of contributions, which at the same time can explain high level of diffuse neutrino flux in 10-100 TeV energy range observed by IceCube \cite{icecube_science,icecube_prl,icecube_cascade20}. Soft spectrum of neutrino signal in this energy range is inconsistent with conventional extragalactic source modelling  \cite{Murase:2015xka}.   Both large scale diffuse neutrino and gamma-ray fluxes can come  from new types of Galactic sources like decaying dark matter (DM) \cite{berezinsky97,kusenko,serpico,neronov18}, cosmic rays injected by nearby recent supernovae  interacting with walls of Local Bubble \cite{superbubble,neronov18,Bouyahiaoui:2018lew,Bouyahiaoui:2020rkf} or from large-scale cosmic ray halo around the Milky Way \cite{taylor_aharonian,Blasi:2019obb}. 

%Although the DM dominates the mass content of all galaxies including the Milky Way, the nature of the DM particles is not known. Particle physics models of the DM span very wide range of possible particle masses, from tiny fractions of electronvolt (e.g. for axion-like particles) up to the "super-heavy" dark matter with masses up to the Grand Unification scale $\sim 10^{16}$~eV \cite{berezinsky97}.  The DM particles with masses above $\sim 10$~TeV could not be produced in collider experiments. Detection of multi-TeV energy particles resulting from DM interactions, including \gr s and neutrinos,   provides the only possibility for exploration of the DM particle mass window $M>10$~TeV. IceCube telescope has reported an astrophysical neutrino signal in the energy range above 30~TeV \cite{icecube_science,icecube_prl} which could possibly have a contributions from both the cosmic ray interactions in the interstellar medium \cite{smirnov,tchernin14,neronov16,neronov16_1,taylor_aharonian} and from the DM decays \cite{kusenko,serpico,neronov18}. Verification of hypotheses of possible DM decay component of the signal would provide an important milestone for the searches of the particle nature of the DM. 

In what follows we explore the sensitivity of LHAASO telescope \cite{lhaaso} for the diffuse \gr\ flux distributed over large angular scales. We compare the sensitivity with the expected levels of the diffuse emission from cosmic ray interactions in the interstellar medium and from the DM decays. We show that LHAASO will provide detailed mapping of the cosmic ray induced \gr\ flux at all Galactic latitudes, in the energy range overlapping with that of IceCube astrophysical neutrino signal \cite{icecube_science}. This will provide an identification of the Galactic component of the astrophysical neutrino flux first predicted by \citet{smirnov}. It will also reveal the characteristic energy at which cosmic rays start to free-stream, rather than diffuse, through the Galactic magnetic field. Finally, diffuse flux measurements will provide up to two orders of magnitude improvement of sensitivity for the search of decaying DM consisting of particles with masses in the PeV-EeV range.

%%%%%%%%%%%%%%%%%%%%%%%%%%%%%%%%%%%%%%%%%%
\section{LHAASO sensitivity for diffuse \gr\ flux }
%%%%%%%%%%%%%%%%%%%%%%%%%%%%%%%%%%%%%%%%%%

The main obstacle for the measurements of large scale diffuse \gr\ flux with ground-based \gr\ telescopes is high level of residual charged cosmic ray background.  Contrary to the \gr s coming from isolated point sources, Extensive Air Showers (EAS) produced by diffuse \gr s  could not be distinguished from the EAS produced by  background charged cosmic rays based on directional information. Still the diffuse \gr\ flux varies as a function of the Right Ascension and declination, while the residual charged particle background rate depends mostly on zenith and azimuth angles. This difference provides a possibility for the measurements of the diffuse \gr\ flux even in the presence of much stronger charged cosmic ray background.   

Imaging Atmospheric Cherenkov telescopes (IACT) can suppress the charged particle background  down to the "minimal possible" level of the cosmic ray electron flux  \cite{kraus,kerszberg,hess_icrc2017,neronov20}. This opens a possibility of the study of the diffuse \gr\ flux in multi-TeV energy range where the cosmic ray electron flux decreases to the level comparable to the diffuse \gr\ flux \cite{neronov20}. 

\begin{figure}
    \includegraphics[width=\linewidth]{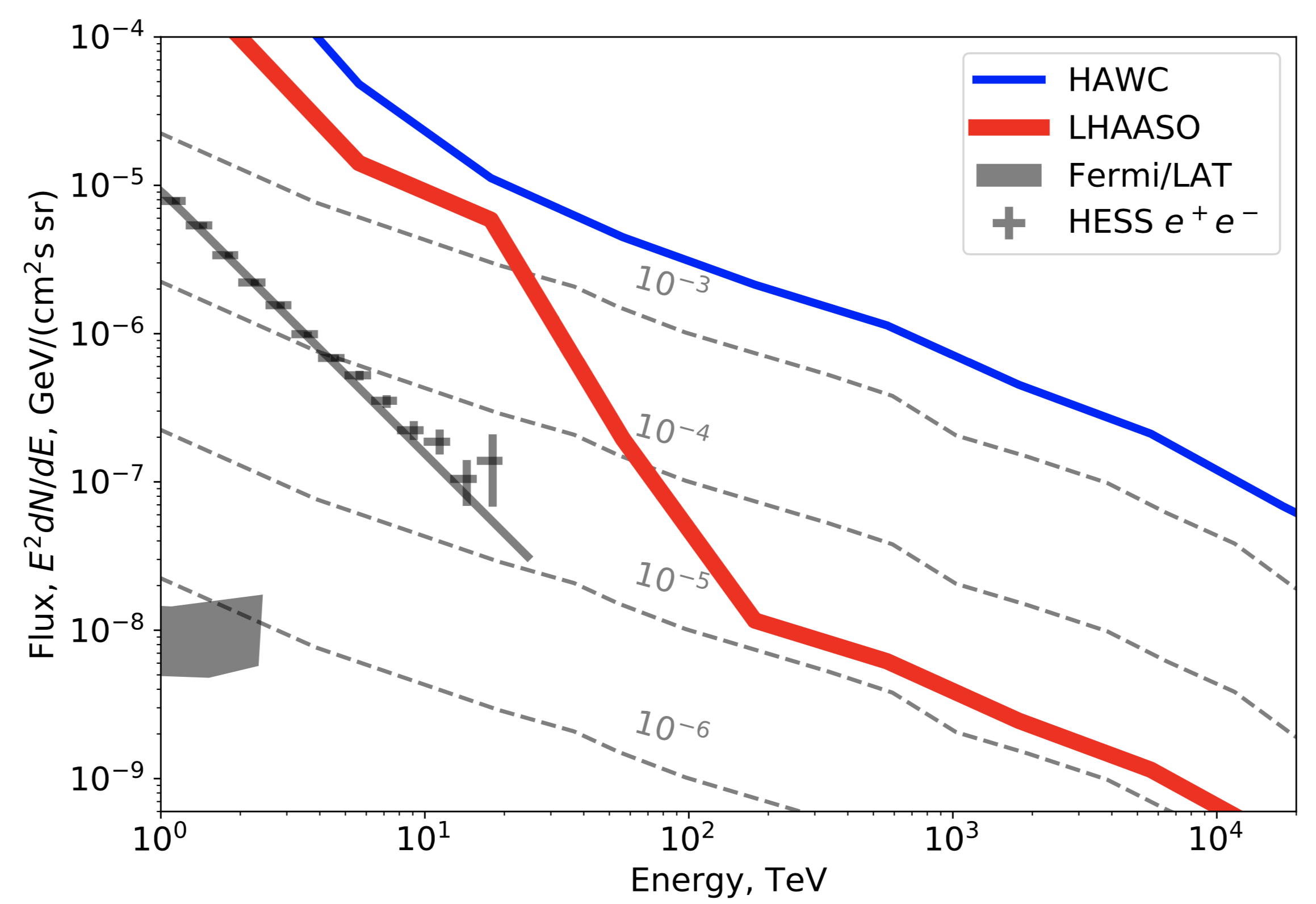}
    \caption{Residual charged cosmic ray backgrounds for the diffuse \gr\ detection in Fermi/LAT \citep{neronov_semikoz19}, HESS (electron spectrum analysis) \citep{hess_icrc2017}, HAWC \citep{hawc} and LHAASO \citep{lhaaso_electrons}. Grey dashed lines with markers show the fractional levels of the overall cosmic ray flux (from $10^{-6}$ to $10^{-3}$).  }
    \label{fig:background}
\end{figure}

This minimal possible charged particle backgorund could not be reached with water Cherenkov detectors such as HAWC \cite{hawc} and Water Cherenkov Detector Array (WCDA) of LHAASO \cite{lhaaso} for which the background suppression techniques  provide moderate efficiency   in the energy range below 10~TeV, based on the imaging of the lateral distribution of particles in the EAS. Comparison of the background levels of HAWC, LHAASO with the "minimal possible" background level in $E<20$~TeV energy range  is shown in Fig. \ref{fig:background}.

%%%%%%%%%%%%%%%%%%%%%%%%%%%%%%%%%%%%%%%%%%
%%%%%%%%%%%%%%%%%%%%%%%%%%%%%%%%%%%%%%%%%%

In the energy range above 20 TeV the background rejection performance of LHAASO  rapidly improves due to the possibility of  detection of the muon component of the EAS with the km2a array \cite{lhaaso,lhaaso_electrons}. The level of the residual background of cosmic ray nuclei (protons) achieved with these technique reaches $\sim 10^{-5}$ of the cosmic ray flux in the energy range $E\sim 100$~TeV (Fig. \ref{fig:background}). 

%%%%%%%%%%%%%%%%%%%%%%%%%%%%%%%%%%%%%%%%%%
\begin{figure}
    \includegraphics[width=\linewidth]{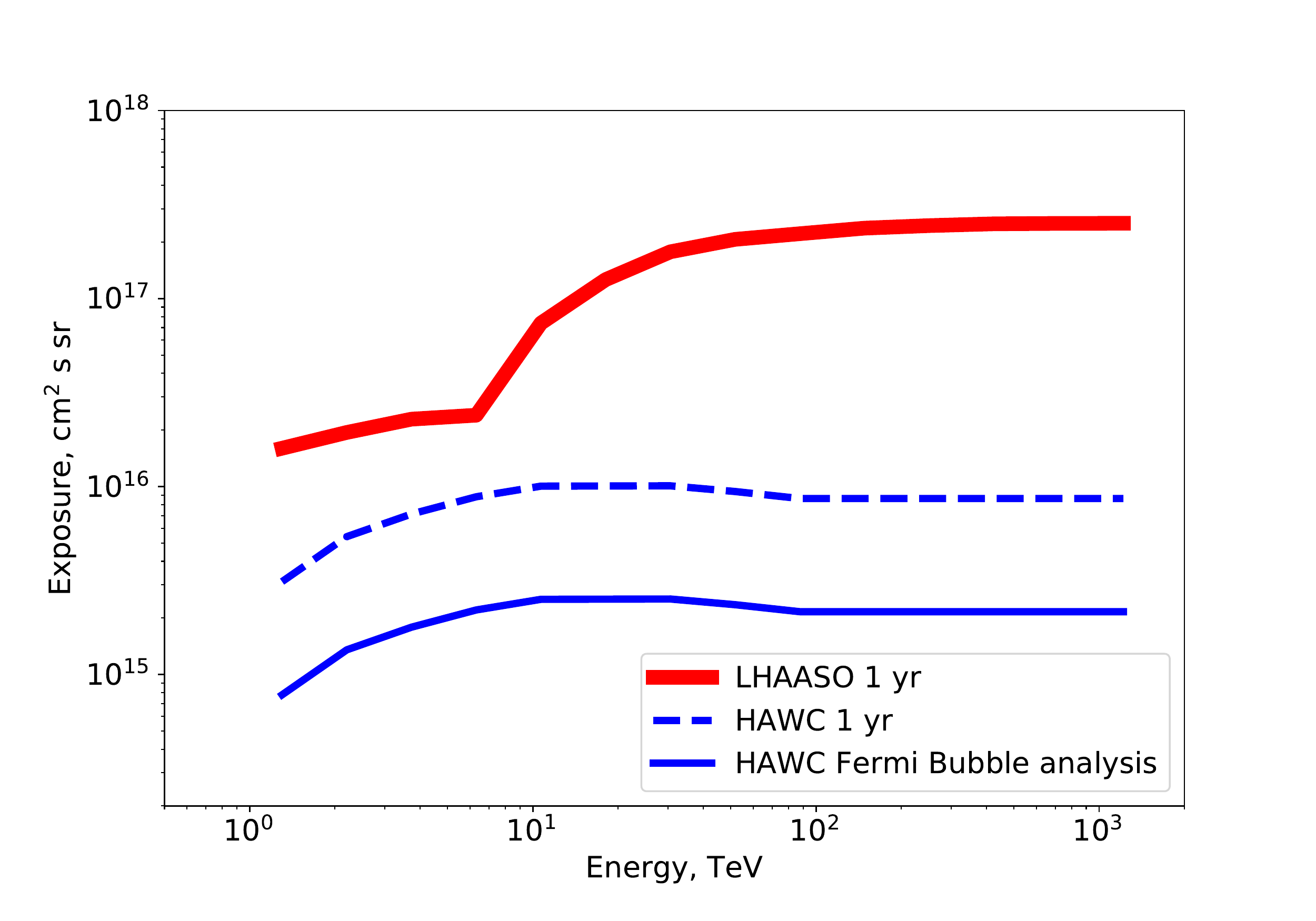}
    \caption{Comparison of one-year exposures of HAWC \citep{hawc} and LHAASO \citep{lhaaso_electrons} of a sky revion within the telescope field-of-view with the HAWC exposure of the Fermi Bubble region considered for the dark matter decay signal search by \citet{hawc_dm,hawc_bubble}. }
    \label{fig:exposure}
\end{figure}
%%%%%%%%%%%%%%%%%%%%%%%%%%%%%%%%%%%%%%%%%%

The level of the residual charged particle background flux $F_B$ determines the sensitivity for the diffuse \gr\ flux from a sky region within the field-of-view of a solid angle $\Omega$ for a telescope with effective collection area $A$  in a given exposure time $T$: the minimal detectable flux should at least be higher than the statistical fluctuations of the background: 
\begin{equation}
\label{eq:one}
F>5\sqrt{F_B/(\Omega  T A)}
\end{equation}
The exposure $\Omega TA$ of LHAASO is compared to that of HAWC in Fig. \ref{fig:exposure}.  The annual exposures are  calculated using the information on the effective collection areas  at zenith angles $\Theta_z<30^\circ$. For comparison we show in the same picture the exposure of the HAWC analysis of the Fermi Bubble region estimated based on the information given in Ref. \cite{hawc_bubble}.

%%%%%%%%%%%%%%%%%%%%%%%%%%%%%%%%%%%%%%%%%%
\begin{figure}
    \includegraphics[width=\linewidth]{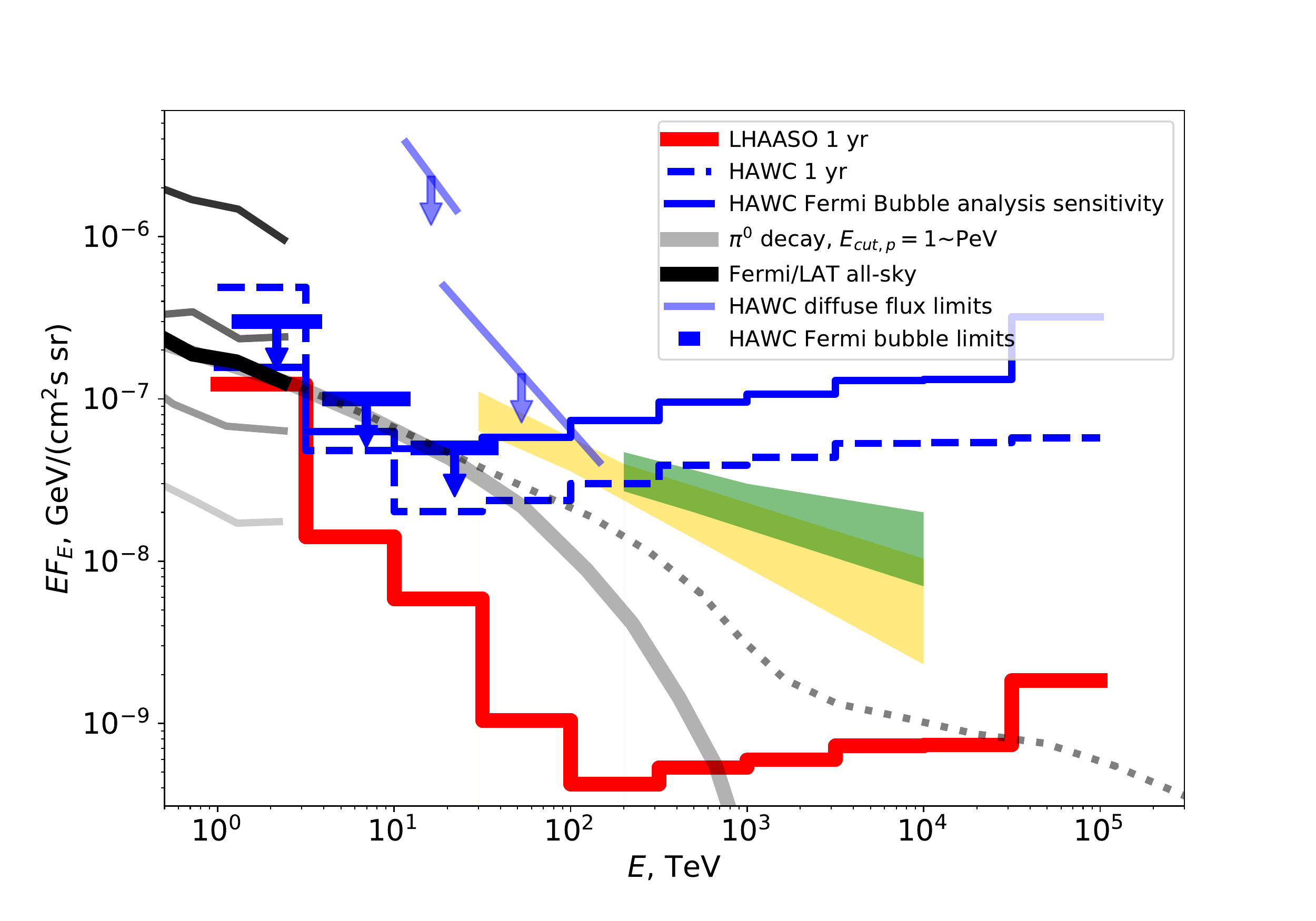}
    \caption{Comparison of the sensitivity LHAASO and HAWC with model predictions of gamma-ray flux from cosmic ray interaction in the interstellar medium. Thin grey thin shaded levels in 0.3-3 TeV energy range show the measurements of diffuse Galactic \gr\ flux with Fermi/LAT, in the sky regions $|l|<30^\circ,\ |b|<2^\circ$; $150^\circ<l<210^\circ,\ |b|<2^\circ$; $10^\circ<|b|<30^\circ$; $|b|>50^\circ$ (from top to bottom), reported by \citet{neronov_semikoz19}. Black thick line shows the all sky averaged flux level \cite{neronov18}. Grey thick line shows a model of pion decay emission produced by proton powerlaw spectrum with cut-off at $E_{p,cut}=1$ PeV.  Grey thick dotted line shows the spectrum without a high-energy cut-off but modified by the effect of $\gamma\gamma$ pair production on Cosmic Microwave Background photons. Yellow and green butterflies show the measurements of the astrophysical neutrino spectrum by IceCube \cite{icrc2019,icecube_cascade20}. HAWC limits on the flux from Fermi Bubble region are from Ref. \cite{hawc_bubble}. Limits on diffuse flux are from \cite{hawc_diffuse}.}
    \label{fig:CR}
\end{figure}
%%%%%%%%%%%%%%%%%%%%%%%%%%%%%%%%%%%%%%%%%%

To calculate  the sensitivity  for the all-sky diffuse \gr\ flux we follow standard approach for the differential sensitivity estimate in \gr\ astronomy. We calculate the minimal detectable flux in individual energy bins (we choose the energy binning homogeneous in logarithm of energy, with two bins per decade, given moderate energy resolution of the water Cherenkov detectors). Apart from relation (\ref{eq:one}), we require that the detectable flux should be at least larger than  $10^{-3}$ of the residual charged particle background, i.e. higher than the flux levels at which dipole anisotropy of the cosmic ray background is detected \cite{cr_anisotropy}.  We also require that  the flux should be high enough to produce at least 10 event counts in a given exposure. 

The resulting sensitivity is shown in Fig. \ref{fig:CR}. We have verified our sensitivity calculation via comparison of the estimate of sensitivity obtained with the method described above for HAWC Fermi Bubble exposure with the results on flux upper limits reported by HAWC  \citet{hawc_bubble} we find a good agreement. This comparison is shown in Fig. \ref{fig:CR}. A discrepancy at the lowest energy is due to the face what we adopt a conservative assumption that the minimal detectable flux is at the level of the dipole anisotropy of the residual cosmic ray background, while the analysis of Ref. \cite{hawc_bubble} reaches lower level via dedicated modelling of the dipole and smaller scale anisotropies.

%%%%%%%%%%%%%%%%%%%%%%%%%%%%%%%%%%%%%%%%%%
\section{\gr\ signal from cosmic ray interactions}
%%%%%%%%%%%%%%%%%%%%%%%%%%%%%%%%%%%%%%%%%%

From Fig. \ref{fig:CR} one can see that the sensitivity limit of LHAASO is well below the expected level of diffuse \gr\ flux from the sky in $E\gtrsim 10$ TeV energy range \cite{neronov18,neronov_semikoz19} in all sky segments. Only in the high Galactic latitude regions, $|b|>50^\circ$ the diffuse sky flux could possibly be marginally detectable by LHAASO, if its spectrum extends as a powerlaw to 10-100 TeV energy range \cite{neronov_semikoz19}. 

The cosmic ray spectrum in the local Galaxy has a pronounced "knee" softening feature of unknown origin 
at energy around 3 PeV in all particle spectrum (see discussion of both observations and interpretations of knee in recent review  \cite{review_CR}).   It is possible that the spectral softening at the knee is due to the change of regime of propagation of cosmic rays in the interstellar medium \cite{1971CoASP...3..155S,1993A&A...268..726P,Candia:2002we}. Lower energy cosmic rays are efficiently scattered off inhomogeneities of the turbulent component of Galactic magnetic field, while higher energy cosmic rays with Larmor radius above maximum scale of turbulent field stream along the ordered magnetic field lines \cite{Giacinti:2014xya,Giacinti:2015hva,Giacinti:2017dgt}. The exact energy of such regime change depends on the structure of magnetic field \cite{jansson}. The Galactic magnetic field varies across the Galactic disk and the energy of the knee feature should therefore also vary. It is, however, not possible to observe such variability with the direct cosmic ray measurements which are available only on the Earth location.  

Interactions of cosmic rays with energies in the PeV range result in production of \gr s with energies in the 10-100 TeV range. A feature in the PeV cosmic ray flux induces a feature in the diffuse \gr\ flux. Therefore, measurements of the 10-100 TeV diffuse \gr\ flux from different parts of the Galaxy provide a possibility to measure the position of the knee of cosmic ray spectra at different locations of the Galaxy. Comparison of LHAASO sensitivity with the sky-average flux model of pion decay emission generated by a cut-off powerlaw distribution of protons with cut-off energy $E_{p,cut}=1$~PeV, calculated based on the parameterisations of Ref. \cite{kelner06}, is shown in Fig. \ref{fig:CR}. LHAASO sensitivity for the diffuse \gr\ flux is largely sufficient for detection and mapping of the position of high-energy suppression of the \gr\ flux induced by the knees of cosmic ray spectra at different locations all along the Galactic Plane. This is clear from the comparison of the sky averaged flux measurement by Fermi/LAT in the TeV range (shown by the thick black line) with the diffuse emission flux levels in the inner and outer Galactic Plane regions \cite{neronov_semikoz19} shown by two top thin grey lines in the TeV energy range in Fig. \ref{fig:CR}. 

An alternative model for the origin of the knee is that it represents high-energy cut-off in the injection spectrum of Galactic cosmic rays from dominant component of cosmic ray sources \cite{Stanev:1993tx,Kobayakawa:2000nq,Hillas_2005,Zatsepin:2006ci}. If there are no sources in the Galaxy able to accelerate protons to the energies much above PeV, the Galactic component of the cosmic ray flux would have a high-energy cut-off. Given that the escape time of PeV cosmic rays from the Galaxy is relatively short, only a small number of individual cosmic ray sources contributes to the cosmic ray content of the Galaxy in the PeV range at any given moment of time.  In particular, only one nearby source can dominate cosmic ray flux around knee \cite{Erlykin:1997bs,Erlykin:2000jm,Erlykin2015}. Such source could be e.g. Vela supernova, which also can be responsible for large fraction of diffuse neutrino flux \cite{Bouyahiaoui:2018lew,Bouyahiaoui:2020rkf}.

Similarly to the escape model of the knee, it is not possible to test cut-off model with direct cosmic ray measurements. The test is possible only with  \gr\ observations. Contrary to the escape model of the knee, variations of the knee positions across the Galactic disk are not expected to correlate with the variations of the structure of Galactic magnetic field in this "source spectral cut-off" model. This provides a possibility for the test of both models with LHAASO observations in different sky directions.  

It is possible that some Galactic cosmic ray sources produce cosmic rays with energies well above the knee. In this case one expects to observe the diffuse emission spectrum without a high-energy cut-off in the 10-100 TeV range from a sky region around such sources. This possibility is shown by the thick dotted grey line in Fig. \ref{fig:CR}.  Even in the absence of the cut-off in the emission spectrum, the spectrum of \gr s from a source in the Galactic Centre region is expected to show strong deviation from a powerlaw. The "dip" spectral feature visible in this spectrum in the PeV energy range is due to the effect of $\gamma\gamma$ pair production on Cosmic Microwave Background photons \cite{gould}.  This type of features in the spectra of isolated sources and diffuse emission is also detectable by LHAASO.  

%%%%%%%%%%%%%%%%%%%%%%%%%%%%%%%%%%%%%%%%%%
\begin{figure}
    \includegraphics[width=\linewidth]{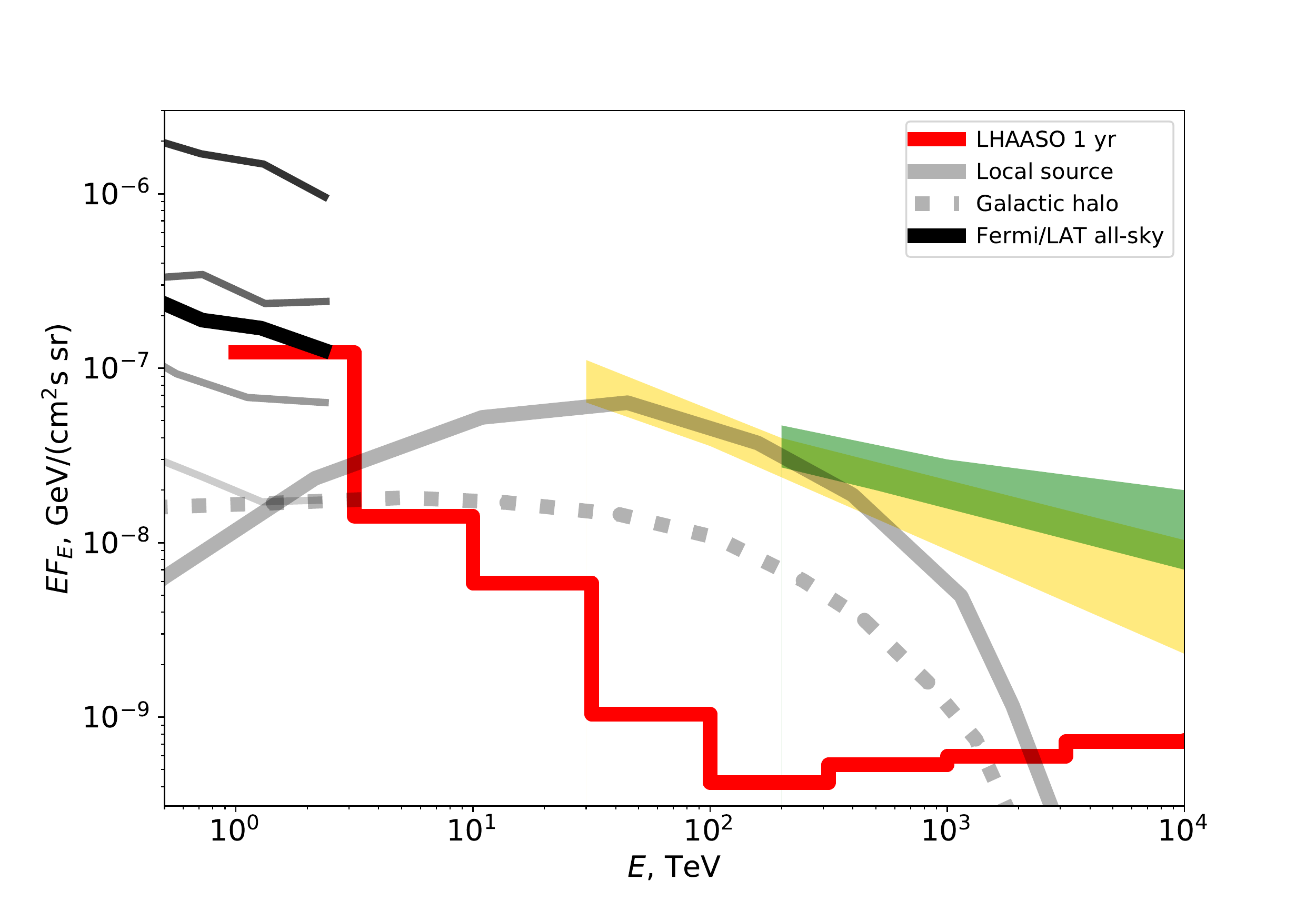}
    \caption{Comparison of the sensitivity LHAASO and HAWC with predictions of the local source (thick grey solid line) and large-scale cosmic ray halo (thick dotted grey line) models. Thin grey shaded levels in 0.3-3 TeV energy range show the measurements of diffuse Galactic \gr\ flux with Fermi/LAT, in the sky regions $|l|<30^\circ,\ |b|<2^\circ$; $150^\circ<l<210^\circ,\ |b|<2^\circ$; $10^\circ<|b|<30^\circ$; $|b|>50^\circ$ (from top to bottom), reported by \citet{neronov_semikoz19}. Thick black line is the sky average diffuse flux measurement \cite{neronov18}. Yellow and green butterflies show the measurements of the astrophysical neutrino spectrum by IceCube \cite{icrc2019}. }
    \label{fig:CR1}
\end{figure}
%%%%%%%%%%%%%%%%%%%%%%%%%%%%%%%%%%%%%%%%%%

The diffuse \gr\ flux from cosmic ray interactions is accompanied by the neutrino flux with comparable spectral characteristics. Therefore, measurement of the diffuse \gr\  flux from all over the sky in the energy band 10 TeV - 10 PeV by LHAASO will "nail" down" the Galactic part of the astrophysical neutrino signal found by IceCube \cite{icecube_science,icecube_prl,icecube_cascade20} (shown in Fig. \ref{fig:CR}), thus providing at least a partial resolution of the problem of the origin of astrophysical neutrinos. 

The overall anisotropy of the astrophysical neutrino flux does not reveal strong excess toward the Galactic Plane or the Galactic Centre direction. This suggests that either the Galactic component does not dominate the astrophysical neutrino flux or that there is a new Galactic flux component which appears in the Multi-TeV energy range and is distributed all over the sky, rather than concentrated toward the Galactic Plane. Such new component could be due to interactions of cosmic rays from nearby sources  \cite{neronov18,Bouyahiaoui:2018lew,Bouyahiaoui:2020rkf} or emission from the large scale cosmic ray halo around the Milky Way \cite{taylor_aharonian,Blasi:2019obb}. Predictions for \gr\ fluxes in these models are shown in Fig. \ref{fig:CR1}. Both the local source and the large scale halo fluxes are bound to be at most at the level of the high Galactic latitude \gr\ flux in the TeV energy range. 

The large scale halo cosmic ray spectrum is close to the $E^{-2}$ powerlaw which is determined by the injection spectrum of cosmic rays from Galactic sources. The halo spectrum has a cut-off at the energy is the characteristic maximal  energy attainable in the Galactic cosmic ray sources. The \gr\ spectrum of the halo follows the powerlaw of the parent proton spectrum (close to $E^{-2}$ and has a cut-off at the energy by a factor $\sim 30$ below the cut-off energy of the parent proton spectrum, because the characteristic energy of the pion decay \gr s is much below the energy of the parent protons \cite{kelner06}. 

The spectrum diffuse emission from the cosmic ray halo around a local source is harder than $E^{-2}$ because low energy cosmic rays are still retained in the source region and could not escape to the interstellar medium. As a result, the \gr\ and neutrino flux levels could reach the level of the IceCube neutrino flux in the 100 TeV energy range. The local source spectrum also has a cut-off at the energy determined by the maximal energy of cosmic rays accelerated in the source \cite{neronov18,Bouyahiaoui:2018lew,Bouyahiaoui:2020rkf}.

%%%%%%%%%%%%%%%%%%%%%%%%%%%%%%%%%%%%%%%%%%
\section{\gr\ signal from the decaying dark matter}
%%%%%%%%%%%%%%%%%%%%%%%%%%%%%%%%%%%%%%%%%%

Cosmic ray interactions in the interstellar medium provide a guaranteed source of neutrinos and \gr s with energy range above 10 TeV. Apart from this guaranteed source, other "unexpected" sources might appear on the sky, like the DM decay signal. It has unique spectral and imaging properties and could be readily distinguished from the the diffuse flux from cosmic ray interactions. 
 
The best strategy for the indirect search of the decaying DM is best performed with telescopes providing the largest "grasp" $G=A\Omega$ \cite{boyarsky06,boyarsky07,boyarsky08}. HAWC and LHAASO are the detectors with the largest grasp in the very-high-energy \gr\ band and are therefore well suited for the DM search.

The signal from the Galactic DM halo is 
\begin{equation}
\label{eq:f}
    \frac{dF_{DM}}{d\Omega}=\frac{\kappa\Gamma_{DM}}{4\pi}\int_{los}\rho_{DM}(r)dl
\end{equation}
where $\rho_{DM}(r)$ is the DM density as a function of the radius from the Galactic Center, $\kappa$ is the fraction of the rest energy of the DM particles transferred to \gr s and $\Gamma_{DM}=1/\tau_{DM}$ is the decay width which is inverse of the DM decay time $\tau_{DM}$. Typical variations of the DM column density $\int_{los}\rho_{DM}dl$ across the sky directions are within a factor of $2$ around the sky-average value. In our estimates we use the DM column density calculation for the Navarro-Frenk-White (NFW) density profile
\begin{equation}
    \rho_{DM}=\frac{\rho_0}{(r/r_0)(1+r/r_0)^2}
\end{equation}
with $\rho_0=0.2$~GeV/cm$^3$ and core radius $r_0=21.5$~kpc \cite{klypin,boyarsky06}.

%%%%%%%%%%%%%%%%%%%%%%%%%%%%%%%%%%%%%%%%%%
\begin{figure}
    \includegraphics[width=\linewidth]{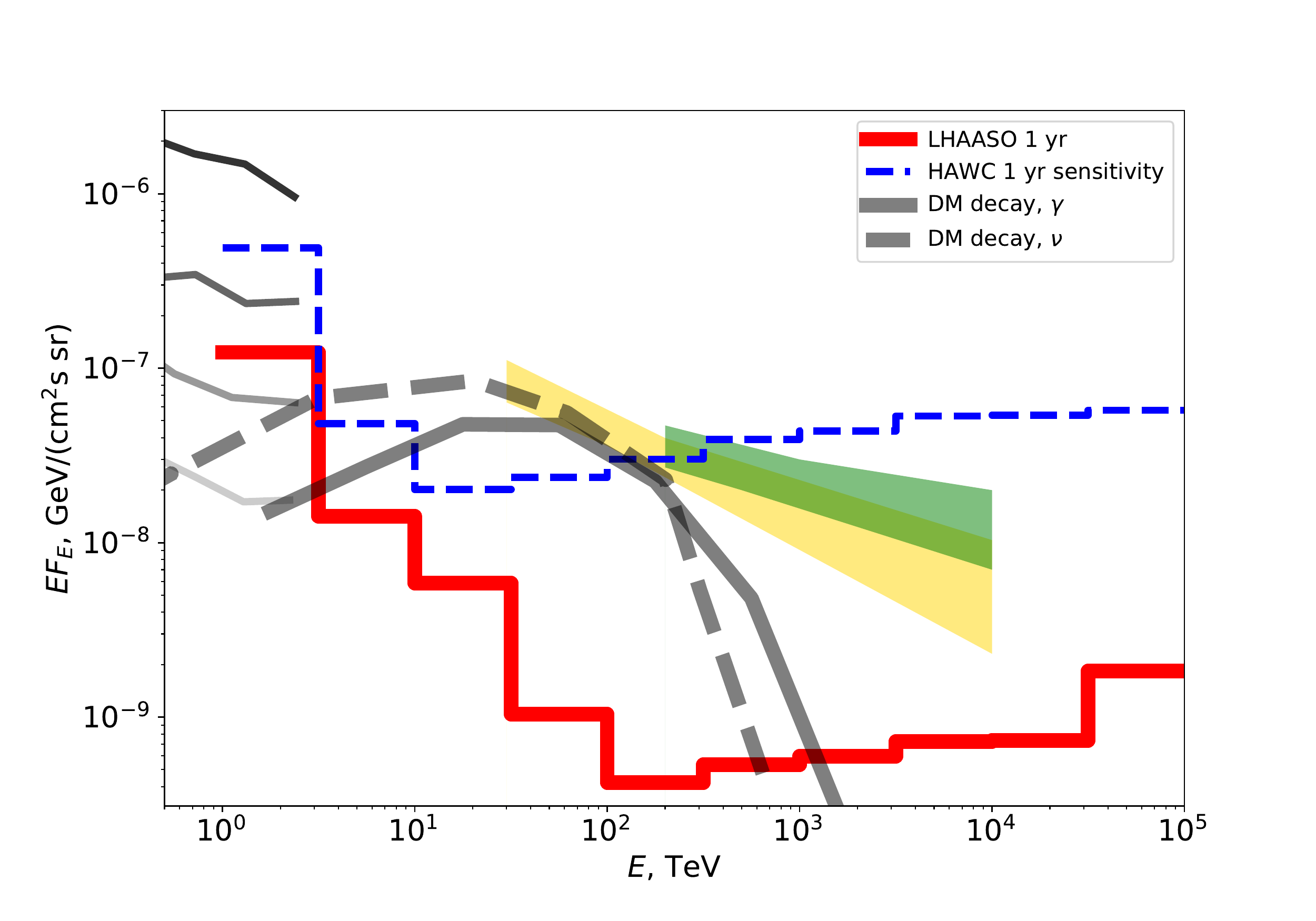}
    \caption{Comparison of the sensitivity LHAASO and HAWC with the expected sky-averaged multi-messenger signal from decaying DM with $\tau_{DM}=3\times 10^{27}$~s, and DM particle mass $m_{DM}=5$~PeV, which could explain the IceCube astrophysical neutrino flux \cite{neronov18}. Thin grey shaded levels in 0.3-3 TeV energy range show the measurements of diffuse Galactic \gr\ flux with Fermi/LAT, in the sky regions $|l|<30^\circ,\ |b|<2^\circ$; $150^\circ<l<210^\circ,\ |b|<2^\circ$; $10^\circ<|b|<30^\circ$; $|b|>50^\circ$ (from top to bottom), reported by \citet{neronov_semikoz19}.}
    \label{fig:DM}
\end{figure}
%%%%%%%%%%%%%%%%%%%%%%%%%%%%%%%%%%%%%%%%%%

The cosmological contribution to the DM decay signal is suppressed in the \gr\ band due to the pair production of the photons of the Extragalactic Background Light (EBL) \cite{franceschini08}. This is not the case for the neutrino signal, which still has the cosmological (isotropic) component. This component reduces the scale of variations of the signal across the sky in the neutrino channel.

Different strategies for the search of the DM decay signal are possible. HAWC analysis \cite{hawc_dm} has adopted an approach in which stronger signal from the directions around the Galactic Center (more precisely, the region of Fermi Bubble) is searched and the rest of the sky is considered for the background estimate. An alternative possibility is to search for somewhat weaker (by a factor of two, on average) signal across the entire sky. An advantage of the latter approach is larger exposure available for the full-sky search. Assuming that the Fermi Bubble region analysed by \citet{hawc_dm} spanned an angle $\Theta_{FB}\lesssim 0.5$~sr, while the full sky available for HAWC is a strip within declination range from $-25^\circ$ to $65^\circ$, with total $\Omega\simeq 7.3$~sr, one finds that the exposure of the Fermi Bubble region is a fraction $\Omega_{FB}/\Omega\simeq 7\%$ of the HAWC annual exposure. The signal-to-noise ratio scales as a square root of time, and the full-sky exposure would provide an increase of the DM signal-to-noise ratio by a factor of $\simeq 2$ on one-year observation time span, compared to the Fermi Bubble region exposure, in spite of the lower average flux.

%%%%%%%%%%%%%%%%%%%%%%%%%%%%%%%%%%%%%%%%%%
\begin{figure}
    \includegraphics[width=\linewidth]{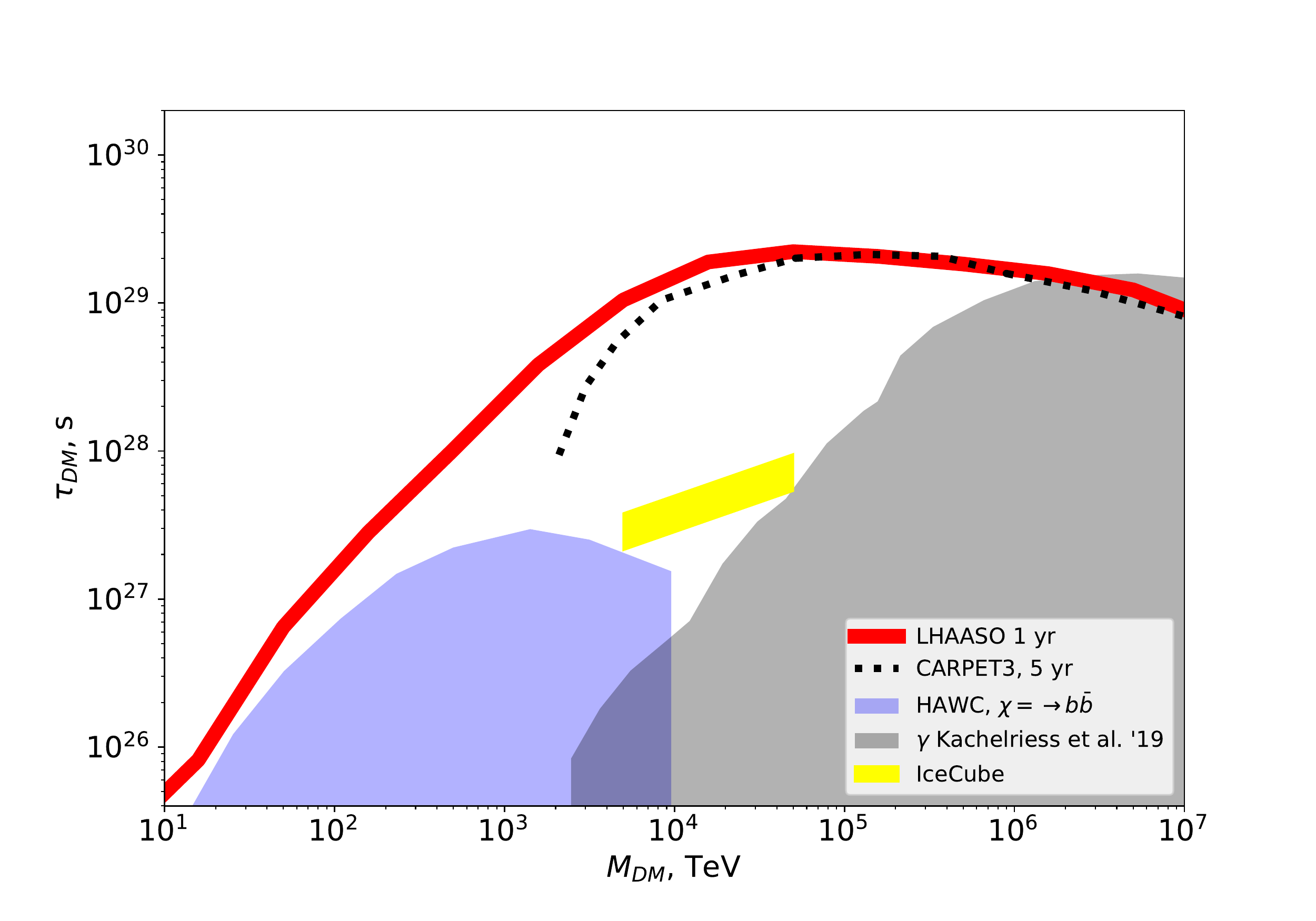}
    \caption{Sensitivity of LHAASO for the measurement of dark matter decay time (for  DM decaying into quarks). Yellow band shows the range of decay times for which DM decays give sizeable contribution to the IceCube neutrino signal \citep{neronov18}. Blue and grey shaded regions show the existing bounds imposed by HAWC \cite{hawc_dm} and ultra-high-energy cosmic ray experiments \cite{kalashev}.  and dashed curves are from the HAWC search of the DM decay signal in the Fermi Bubble regions \citep{hawc_dm}.  }
    \label{fig:tau}
\end{figure}
%%%%%%%%%%%%%%%%%%%%%%%%%%%%%%%%%%%%%%%%%%

Use of the full-sky exposure, rather than of limited sky region around the Galactic Centre is important also in the view of uncertainties of the Galactic diffuse \gr\ emission unrelated to the DM decay flux. This Galactic diffuse emission provides a background on top of which the DM decay signal is detected. Even though this background is possibly sub-dominant compared to the residual charged particle background in \gr\ telescopes, it might still be stronger than the DM decay signal. 

Fig. \ref{fig:DM} shows a comparison of sensitivity of LHAASO with the model predictions for the \gr\ and neutrino fluxes for the DM with mass $m_{DM}=5$~PeV decaying into quarks on the time scale $\tau_{DM}=3\times 10^{27}$~s \cite{neronov18}. The decay time is chosen so that the neutrino flux level is comparable to the IceCube astrophysical neutrino flux. From this figure one can see that  LHAASO sensitivity for the diffuse \gr\ flux will be sufficient for the detection of the \gr\ signal in such DM decay model.

To estimate LHAASO sensitivity reach for the decaying DM search, we follow the approach of Ref. \cite{hawc_dm} to estimate the significance of detection of the DM decay signal sampled from all sky in the field-of-view. In each energy bin we compare the DM decay flux levels for different values of $m_{DM},\tau_{DM}$ with the residual charged particle background levels and calculate by how much is  the $\chi^2$ of the fit of the signal+background data is inconsistent with the backgorund-only model in all energy bins. In this way we find the minimal detectable DM decay flux as a function of the DM mass for the model of Ref. \cite{neronov18} of DM decaying into quark-antiquark pair. We then convert the estimate of the minimal detectable flux into the estimate of the maximal measureable DM decay time using Eq. (\ref{eq:f}). The result is shown in Fig. \ref{fig:tau}. From this figure one could see that LHAASO will explore the range of DM decay times up to $\tau_{DM}\sim 3\times 10^{29}$~s over a wide DM mass range $m_{DM}>10$~PeV. In the mass range $10$~TeV$<m_{DM}<10$~PeV LHAASO will provide a factor of 3-to-10 improvement of sensitivity compared to HAWC. In any case, LHAASO will fully test a model in which non-negligible fraction of the astrophysical neutrino flux is generated by the DM decays. Comparing LHAASO sensitivity with the limits imposed by the non-detection of \gr\ signal by ultra-high-energy cosmic ray experiments \cite{kalashev} we find that the LHAASO will mostly provide better sensitivity in the DM mass range below EeV. 

%%%%%%%%%%%%%%%%%
\section{Conclusions}
%%%%%%%%%%%%%%%%%%

We have explored the potential of LHAASO for the study of diffuse \gr\ emission signals in the TeV - PeV energy range. We find that its sensitivity will be largely sufficient for the measurement of Galactic diffuse \gr\ flux generated by interactions of cosmic rays with energies up to the knee (Figs. \ref{fig:CR}, \ref{fig:CR1}). LHAASO study of the diffuse \gr\ flux will provide a clue for solution of the problem of the nature of the knee feature of the cosmic ray spectrum. It will also nail down the Galactic component of the astrophysical neutrino flux, including possible contributions from Local Bubble and Galactic halo. 
LHAASO will provide a major improvement of sensitivity for the search of the \gr\ signal from decaying heavy DM particles with masses in the TeV-EeV range (Fig. \ref{fig:tau}). This improvement will be sufficient for the full test of a model of the IceCube astrophysical neutrino signal in which a sizeable fraction of the neutrino flux originates from DM decays in the Galactic Halo and in the distant Universe.   

\bibliography{LHAASO,Diffuse_gamma_HESS}

\begin{thebibliography}{56}
\expandafter\ifx\csname natexlab\endcsname\relax\def\natexlab#1{#1}\fi
\expandafter\ifx\csname bibnamefont\endcsname\relax
  \def\bibnamefont#1{#1}\fi
\expandafter\ifx\csname bibfnamefont\endcsname\relax
  \def\bibfnamefont#1{#1}\fi
\expandafter\ifx\csname citenamefont\endcsname\relax
  \def\citenamefont#1{#1}\fi
\expandafter\ifx\csname url\endcsname\relax
  \def\url#1{\texttt{#1}}\fi
\expandafter\ifx\csname urlprefix\endcsname\relax\def\urlprefix{URL }\fi
\providecommand{\bibinfo}[2]{#2}
\providecommand{\eprint}[2][]{\url{#2}}

\bibitem[{\citenamefont{{Ackermann} et~al.}(2012)\citenamefont{{Ackermann},
  {Ajello}, {Atwood}, {Baldini}, {Ballet}, {Barbiellini}, {Bastieri},
  {Bechtol}, {Bellazzini}, {Berenji} et~al.}}]{fermi_diffuse_2012}
\bibinfo{author}{\bibfnamefont{M.}~\bibnamefont{{Ackermann}}},
  \bibinfo{author}{\bibfnamefont{M.}~\bibnamefont{{Ajello}}},
  \bibinfo{author}{\bibfnamefont{W.~B.} \bibnamefont{{Atwood}}},
  \bibinfo{author}{\bibfnamefont{L.}~\bibnamefont{{Baldini}}},
  \bibinfo{author}{\bibfnamefont{J.}~\bibnamefont{{Ballet}}},
  \bibinfo{author}{\bibfnamefont{G.}~\bibnamefont{{Barbiellini}}},
  \bibinfo{author}{\bibfnamefont{D.}~\bibnamefont{{Bastieri}}},
  \bibinfo{author}{\bibfnamefont{K.}~\bibnamefont{{Bechtol}}},
  \bibinfo{author}{\bibfnamefont{R.}~\bibnamefont{{Bellazzini}}},
  \bibinfo{author}{\bibfnamefont{B.}~\bibnamefont{{Berenji}}},
  \bibnamefont{et~al.}, \bibinfo{journal}{\apj} \textbf{\bibinfo{volume}{750}},
  \bibinfo{eid}{3} (\bibinfo{year}{2012}), \eprint{1202.4039}.

\bibitem[{\citenamefont{{Neronov} and {Semikoz}}(2019)}]{neronov_semikoz19}
\bibinfo{author}{\bibfnamefont{A.}~\bibnamefont{{Neronov}}} \bibnamefont{and}
  \bibinfo{author}{\bibfnamefont{D.~V.} \bibnamefont{{Semikoz}}},
  \bibinfo{journal}{arXiv e-prints} \bibinfo{eid}{arXiv:1907.06061}
  (\bibinfo{year}{2019}), \eprint{1907.06061}.

\bibitem[{\citenamefont{{Abramowski} et~al.}(2014)\citenamefont{{Abramowski},
  {Aharonian}, {Ait Benkhali}, {Akhperjanian}, {Ang{\"u}ner}, {Backes},
  {Balenderan}, {Balzer}, {Barnacka}, {Becherini} et~al.}}]{hess_diffuse}
\bibinfo{author}{\bibfnamefont{A.}~\bibnamefont{{Abramowski}}},
  \bibinfo{author}{\bibfnamefont{F.}~\bibnamefont{{Aharonian}}},
  \bibinfo{author}{\bibfnamefont{F.}~\bibnamefont{{Ait Benkhali}}},
  \bibinfo{author}{\bibfnamefont{A.~G.} \bibnamefont{{Akhperjanian}}},
  \bibinfo{author}{\bibfnamefont{E.~O.} \bibnamefont{{Ang{\"u}ner}}},
  \bibinfo{author}{\bibfnamefont{M.}~\bibnamefont{{Backes}}},
  \bibinfo{author}{\bibfnamefont{S.}~\bibnamefont{{Balenderan}}},
  \bibinfo{author}{\bibfnamefont{A.}~\bibnamefont{{Balzer}}},
  \bibinfo{author}{\bibfnamefont{A.}~\bibnamefont{{Barnacka}}},
  \bibinfo{author}{\bibfnamefont{Y.}~\bibnamefont{{Becherini}}},
  \bibnamefont{et~al.}, \bibinfo{journal}{\prd} \textbf{\bibinfo{volume}{90}},
  \bibinfo{eid}{122007} (\bibinfo{year}{2014}), \eprint{1411.7568}.

\bibitem[{\citenamefont{{Atwood} et~al.}(2009)\citenamefont{{Atwood}, {Abdo},
  {Ackermann}, {Althouse}, {Anderson}, {Axelsson}, {Baldini}, {Ballet}, {Band},
  {Barbiellini} et~al.}}]{atwood09}
\bibinfo{author}{\bibfnamefont{W.~B.} \bibnamefont{{Atwood}}},
  \bibinfo{author}{\bibfnamefont{A.~A.} \bibnamefont{{Abdo}}},
  \bibinfo{author}{\bibfnamefont{M.}~\bibnamefont{{Ackermann}}},
  \bibinfo{author}{\bibfnamefont{W.}~\bibnamefont{{Althouse}}},
  \bibinfo{author}{\bibfnamefont{B.}~\bibnamefont{{Anderson}}},
  \bibinfo{author}{\bibfnamefont{M.}~\bibnamefont{{Axelsson}}},
  \bibinfo{author}{\bibfnamefont{L.}~\bibnamefont{{Baldini}}},
  \bibinfo{author}{\bibfnamefont{J.}~\bibnamefont{{Ballet}}},
  \bibinfo{author}{\bibfnamefont{D.~L.} \bibnamefont{{Band}}},
  \bibinfo{author}{\bibfnamefont{G.}~\bibnamefont{{Barbiellini}}},
  \bibnamefont{et~al.}, \bibinfo{journal}{\apj} \textbf{\bibinfo{volume}{697}},
  \bibinfo{pages}{1071} (\bibinfo{year}{2009}), \eprint{0902.1089}.

\bibitem[{\citenamefont{{Gould} and {Schr{\'e}der}}(1966)}]{gould}
\bibinfo{author}{\bibfnamefont{R.~J.} \bibnamefont{{Gould}}} \bibnamefont{and}
  \bibinfo{author}{\bibfnamefont{G.}~\bibnamefont{{Schr{\'e}der}}},
  \bibinfo{journal}{\prl} \textbf{\bibinfo{volume}{16}}, \bibinfo{pages}{252}
  (\bibinfo{year}{1966}).

\bibitem[{\citenamefont{{Franceschini}
  et~al.}(2008)\citenamefont{{Franceschini}, {Rodighiero}, and
  {Vaccari}}}]{franceschini08}
\bibinfo{author}{\bibfnamefont{A.}~\bibnamefont{{Franceschini}}},
  \bibinfo{author}{\bibfnamefont{G.}~\bibnamefont{{Rodighiero}}},
  \bibnamefont{and}
  \bibinfo{author}{\bibfnamefont{M.}~\bibnamefont{{Vaccari}}},
  \bibinfo{journal}{\aap} \textbf{\bibinfo{volume}{487}}, \bibinfo{pages}{837}
  (\bibinfo{year}{2008}), \eprint{0805.1841}.

\bibitem[{\citenamefont{{Blumenthal} and {Gould}}(1970)}]{blumenthal}
\bibinfo{author}{\bibfnamefont{G.~R.} \bibnamefont{{Blumenthal}}}
  \bibnamefont{and} \bibinfo{author}{\bibfnamefont{R.~J.}
  \bibnamefont{{Gould}}}, \bibinfo{journal}{Reviews of Modern Physics}
  \textbf{\bibinfo{volume}{42}}, \bibinfo{pages}{237} (\bibinfo{year}{1970}).

\bibitem[{\citenamefont{{Schlickeiser}}(1979)}]{schlickeiser}
\bibinfo{author}{\bibfnamefont{R.}~\bibnamefont{{Schlickeiser}}},
  \bibinfo{journal}{\apj} \textbf{\bibinfo{volume}{233}}, \bibinfo{pages}{294}
  (\bibinfo{year}{1979}).

\bibitem[{\citenamefont{{H.~E.~S.~S. Collaboration}}(2017)}]{hess_icrc2017}
\bibinfo{author}{\bibnamefont{{H.~E.~S.~S. Collaboration}}},
  \bibinfo{journal}{arXiv e-prints} \bibinfo{eid}{arXiv:1709.06442}
  (\bibinfo{year}{2017}), \eprint{1709.06442}.

\bibitem[{\citenamefont{{Kelner} et~al.}(2006)\citenamefont{{Kelner},
  {Aharonian}, and {Bugayov}}}]{kelner06}
\bibinfo{author}{\bibfnamefont{S.~R.} \bibnamefont{{Kelner}}},
  \bibinfo{author}{\bibfnamefont{F.~A.} \bibnamefont{{Aharonian}}},
  \bibnamefont{and} \bibinfo{author}{\bibfnamefont{V.~V.}
  \bibnamefont{{Bugayov}}}, \bibinfo{journal}{\prd}
  \textbf{\bibinfo{volume}{74}}, \bibinfo{eid}{034018} (\bibinfo{year}{2006}),
  \eprint{astro-ph/0606058}.

\bibitem[{\citenamefont{{Kappes} et~al.}(2007)\citenamefont{{Kappes}, {Hinton},
  {Stegmann}, and {Aharonian}}}]{kappes}
\bibinfo{author}{\bibfnamefont{A.}~\bibnamefont{{Kappes}}},
  \bibinfo{author}{\bibfnamefont{J.}~\bibnamefont{{Hinton}}},
  \bibinfo{author}{\bibfnamefont{C.}~\bibnamefont{{Stegmann}}},
  \bibnamefont{and} \bibinfo{author}{\bibfnamefont{F.~A.}
  \bibnamefont{{Aharonian}}}, \bibinfo{journal}{\apj}
  \textbf{\bibinfo{volume}{656}}, \bibinfo{pages}{870} (\bibinfo{year}{2007}),
  \eprint{astro-ph/0607286}.

\bibitem[{\citenamefont{{Berezinskii} and {Smirnov}}(1975)}]{smirnov}
\bibinfo{author}{\bibfnamefont{V.~S.} \bibnamefont{{Berezinskii}}}
  \bibnamefont{and} \bibinfo{author}{\bibfnamefont{A.~I.}
  \bibnamefont{{Smirnov}}}, \bibinfo{journal}{\apss}
  \textbf{\bibinfo{volume}{32}}, \bibinfo{pages}{461} (\bibinfo{year}{1975}).

\bibitem[{\citenamefont{{IceCube Collaboration}}(2013)}]{icecube_science}
\bibinfo{author}{\bibnamefont{{IceCube Collaboration}}},
  \bibinfo{journal}{Science} \textbf{\bibinfo{volume}{342}},
  \bibinfo{eid}{1242856} (\bibinfo{year}{2013}), \eprint{1311.5238}.

\bibitem[{\citenamefont{Aartsen et~al.}(2014)\citenamefont{Aartsen, Ackermann,
  Adams, Aguilar, Ahlers, Ahrens, Altmann, Anderson, Arguelles, Arlen
  et~al.}}]{icecube_prl}
\bibinfo{author}{\bibfnamefont{M.~G.} \bibnamefont{Aartsen}},
  \bibinfo{author}{\bibfnamefont{M.}~\bibnamefont{Ackermann}},
  \bibinfo{author}{\bibfnamefont{J.}~\bibnamefont{Adams}},
  \bibinfo{author}{\bibfnamefont{J.~A.} \bibnamefont{Aguilar}},
  \bibinfo{author}{\bibfnamefont{M.}~\bibnamefont{Ahlers}},
  \bibinfo{author}{\bibfnamefont{M.}~\bibnamefont{Ahrens}},
  \bibinfo{author}{\bibfnamefont{D.}~\bibnamefont{Altmann}},
  \bibinfo{author}{\bibfnamefont{T.}~\bibnamefont{Anderson}},
  \bibinfo{author}{\bibfnamefont{C.}~\bibnamefont{Arguelles}},
  \bibinfo{author}{\bibfnamefont{T.~C.} \bibnamefont{Arlen}},
  \bibnamefont{et~al.} (\bibinfo{collaboration}{IceCube Collaboration}),
  \bibinfo{journal}{Phys. Rev. Lett.} \textbf{\bibinfo{volume}{113}},
  \bibinfo{pages}{101101} (\bibinfo{year}{2014}),
  \urlprefix\url{https://link.aps.org/doi/10.1103/PhysRevLett.113.101101}.

\bibitem[{\citenamefont{{IceCube Collaboration}
  et~al.}(2020)\citenamefont{{IceCube Collaboration}, {Aartsen}, {Ackermann},
  {Adams}, {Aguilar}, {Ahlers}, {Ahrens}, {Alispach}, {Andeen}, {Anderson}
  et~al.}}]{icecube_cascade20}
\bibinfo{author}{\bibnamefont{{IceCube Collaboration}}},
  \bibinfo{author}{\bibfnamefont{M.~G.} \bibnamefont{{Aartsen}}},
  \bibinfo{author}{\bibfnamefont{M.}~\bibnamefont{{Ackermann}}},
  \bibinfo{author}{\bibfnamefont{J.}~\bibnamefont{{Adams}}},
  \bibinfo{author}{\bibfnamefont{J.~A.} \bibnamefont{{Aguilar}}},
  \bibinfo{author}{\bibfnamefont{M.}~\bibnamefont{{Ahlers}}},
  \bibinfo{author}{\bibfnamefont{M.}~\bibnamefont{{Ahrens}}},
  \bibinfo{author}{\bibfnamefont{C.}~\bibnamefont{{Alispach}}},
  \bibinfo{author}{\bibfnamefont{K.}~\bibnamefont{{Andeen}}},
  \bibinfo{author}{\bibfnamefont{T.}~\bibnamefont{{Anderson}}},
  \bibnamefont{et~al.}, \bibinfo{journal}{arXiv e-prints}
  \bibinfo{eid}{arXiv:2001.09520} (\bibinfo{year}{2020}), \eprint{2001.09520}.

\bibitem[{\citenamefont{Murase et~al.}(2016)\citenamefont{Murase, Guetta, and
  Ahlers}}]{Murase:2015xka}
\bibinfo{author}{\bibfnamefont{K.}~\bibnamefont{Murase}},
  \bibinfo{author}{\bibfnamefont{D.}~\bibnamefont{Guetta}}, \bibnamefont{and}
  \bibinfo{author}{\bibfnamefont{M.}~\bibnamefont{Ahlers}},
  \bibinfo{journal}{Phys. Rev. Lett.} \textbf{\bibinfo{volume}{116}},
  \bibinfo{pages}{071101} (\bibinfo{year}{2016}), \eprint{1509.00805}.

\bibitem[{\citenamefont{{Berezinsky} et~al.}(1997)\citenamefont{{Berezinsky},
  {Kachelrie{\ss}}, and {Vilenkin}}}]{berezinsky97}
\bibinfo{author}{\bibfnamefont{V.}~\bibnamefont{{Berezinsky}}},
  \bibinfo{author}{\bibfnamefont{M.}~\bibnamefont{{Kachelrie{\ss}}}},
  \bibnamefont{and}
  \bibinfo{author}{\bibfnamefont{A.}~\bibnamefont{{Vilenkin}}},
  \bibinfo{journal}{\prl} \textbf{\bibinfo{volume}{79}}, \bibinfo{pages}{4302}
  (\bibinfo{year}{1997}), \eprint{astro-ph/9708217}.

\bibitem[{\citenamefont{{Feldstein} et~al.}(2013)\citenamefont{{Feldstein},
  {Kusenko}, {Matsumoto}, and {Yanagida}}}]{kusenko}
\bibinfo{author}{\bibfnamefont{B.}~\bibnamefont{{Feldstein}}},
  \bibinfo{author}{\bibfnamefont{A.}~\bibnamefont{{Kusenko}}},
  \bibinfo{author}{\bibfnamefont{S.}~\bibnamefont{{Matsumoto}}},
  \bibnamefont{and} \bibinfo{author}{\bibfnamefont{T.~T.}
  \bibnamefont{{Yanagida}}}, \bibinfo{journal}{\prd}
  \textbf{\bibinfo{volume}{88}}, \bibinfo{eid}{015004} (\bibinfo{year}{2013}),
  \eprint{1303.7320}.

\bibitem[{\citenamefont{{Esmaili} and {Serpico}}(2013)}]{serpico}
\bibinfo{author}{\bibfnamefont{A.}~\bibnamefont{{Esmaili}}} \bibnamefont{and}
  \bibinfo{author}{\bibfnamefont{P.~D.} \bibnamefont{{Serpico}}},
  \bibinfo{journal}{\jcap} \textbf{\bibinfo{volume}{2013}}, \bibinfo{eid}{054}
  (\bibinfo{year}{2013}), \eprint{1308.1105}.

\bibitem[{\citenamefont{{Neronov} et~al.}(2018)\citenamefont{{Neronov},
  {Kachelrie{\ss}}, and {Semikoz}}}]{neronov18}
\bibinfo{author}{\bibfnamefont{A.}~\bibnamefont{{Neronov}}},
  \bibinfo{author}{\bibfnamefont{M.}~\bibnamefont{{Kachelrie{\ss}}}},
  \bibnamefont{and} \bibinfo{author}{\bibfnamefont{D.~V.}
  \bibnamefont{{Semikoz}}}, \bibinfo{journal}{\prd}
  \textbf{\bibinfo{volume}{98}}, \bibinfo{eid}{023004} (\bibinfo{year}{2018}),
  \eprint{1802.09983}.

\bibitem[{\citenamefont{{Andersen} et~al.}(2018)\citenamefont{{Andersen},
  {Kachelriess}, and {Semikoz}}}]{superbubble}
\bibinfo{author}{\bibfnamefont{K.~J.} \bibnamefont{{Andersen}}},
  \bibinfo{author}{\bibfnamefont{M.}~\bibnamefont{{Kachelriess}}},
  \bibnamefont{and} \bibinfo{author}{\bibfnamefont{D.~V.}
  \bibnamefont{{Semikoz}}}, \bibinfo{journal}{\apjl}
  \textbf{\bibinfo{volume}{861}}, \bibinfo{eid}{L19} (\bibinfo{year}{2018}),
  \eprint{1712.03153}.

\bibitem[{\citenamefont{Bouyahiaoui et~al.}(2019)\citenamefont{Bouyahiaoui,
  Kachelriess, and Semikoz}}]{Bouyahiaoui:2018lew}
\bibinfo{author}{\bibfnamefont{M.}~\bibnamefont{Bouyahiaoui}},
  \bibinfo{author}{\bibfnamefont{M.}~\bibnamefont{Kachelriess}},
  \bibnamefont{and} \bibinfo{author}{\bibfnamefont{D.~V.}
  \bibnamefont{Semikoz}}, \bibinfo{journal}{JCAP}
  \textbf{\bibinfo{volume}{1901}}, \bibinfo{pages}{046} (\bibinfo{year}{2019}),
  \eprint{1812.03522}.

\bibitem[{\citenamefont{Bouyahiaoui et~al.}(2020)\citenamefont{Bouyahiaoui,
  Kachelrieß, and Semikoz}}]{Bouyahiaoui:2020rkf}
\bibinfo{author}{\bibfnamefont{M.}~\bibnamefont{Bouyahiaoui}},
  \bibinfo{author}{\bibfnamefont{M.}~\bibnamefont{Kachelrieß}},
  \bibnamefont{and} \bibinfo{author}{\bibfnamefont{D.~V.}
  \bibnamefont{Semikoz}} (\bibinfo{year}{2020}), \eprint{2001.00768}.

\bibitem[{\citenamefont{{Taylor} et~al.}(2014)\citenamefont{{Taylor}, {Gabici},
  and {Aharonian}}}]{taylor_aharonian}
\bibinfo{author}{\bibfnamefont{A.~M.} \bibnamefont{{Taylor}}},
  \bibinfo{author}{\bibfnamefont{S.}~\bibnamefont{{Gabici}}}, \bibnamefont{and}
  \bibinfo{author}{\bibfnamefont{F.}~\bibnamefont{{Aharonian}}},
  \bibinfo{journal}{\prd} \textbf{\bibinfo{volume}{89}}, \bibinfo{eid}{103003}
  (\bibinfo{year}{2014}), \eprint{1403.3206}.

\bibitem[{\citenamefont{Blasi and Amato}(2019)}]{Blasi:2019obb}
\bibinfo{author}{\bibfnamefont{P.}~\bibnamefont{Blasi}} \bibnamefont{and}
  \bibinfo{author}{\bibfnamefont{E.}~\bibnamefont{Amato}},
  \bibinfo{journal}{Phys. Rev. Lett.} \textbf{\bibinfo{volume}{122}},
  \bibinfo{pages}{051101} (\bibinfo{year}{2019}), \eprint{1901.03609}.

\bibitem[{\citenamefont{{Bai} et~al.}(2019)\citenamefont{{Bai}, {Bi}, {Bi},
  {Cao}, {Chen}, {Chen}, {Chiavassa}, {Cui}, {Dai}, {della Volpe}
  et~al.}}]{lhaaso}
\bibinfo{author}{\bibfnamefont{X.}~\bibnamefont{{Bai}}},
  \bibinfo{author}{\bibfnamefont{B.~Y.} \bibnamefont{{Bi}}},
  \bibinfo{author}{\bibfnamefont{X.~J.} \bibnamefont{{Bi}}},
  \bibinfo{author}{\bibfnamefont{Z.}~\bibnamefont{{Cao}}},
  \bibinfo{author}{\bibfnamefont{S.~Z.} \bibnamefont{{Chen}}},
  \bibinfo{author}{\bibfnamefont{Y.}~\bibnamefont{{Chen}}},
  \bibinfo{author}{\bibfnamefont{A.}~\bibnamefont{{Chiavassa}}},
  \bibinfo{author}{\bibfnamefont{X.~H.} \bibnamefont{{Cui}}},
  \bibinfo{author}{\bibfnamefont{Z.~G.} \bibnamefont{{Dai}}},
  \bibinfo{author}{\bibfnamefont{D.}~\bibnamefont{{della Volpe}}},
  \bibnamefont{et~al.}, \bibinfo{journal}{arXiv e-prints}
  \bibinfo{eid}{arXiv:1905.02773} (\bibinfo{year}{2019}), \eprint{1905.02773}.

\bibitem[{\citenamefont{Kraus}(2018)}]{kraus}
\bibinfo{author}{\bibfnamefont{M.}~\bibnamefont{Kraus}}, Ph.D. thesis,
  \bibinfo{school}{{University of Erlangen}} (\bibinfo{year}{2018}),
  \urlprefix\url{https://ecap.nat.fau.de/wp-content/uploads/2018/07/2018_Kraus_Dissertation.pdf}.

\bibitem[{\citenamefont{Kerszberg}(2017)}]{kerszberg}
\bibinfo{author}{\bibfnamefont{D.}~\bibnamefont{Kerszberg}},
  \bibinfo{type}{Theses}, \bibinfo{school}{{Universit{\'e} Pierre et Marie
  Curie - Paris VI}} (\bibinfo{year}{2017}),
  \urlprefix\url{https://tel.archives-ouvertes.fr/tel-01722819}.

\bibitem[{\citenamefont{{Neronov} and {Semikoz}}(2020)}]{neronov20}
\bibinfo{author}{\bibfnamefont{A.}~\bibnamefont{{Neronov}}} \bibnamefont{and}
  \bibinfo{author}{\bibfnamefont{D.}~\bibnamefont{{Semikoz}}},
  \bibinfo{journal}{arXiv e-prints} \bibinfo{eid}{arXiv:2001.00922}
  (\bibinfo{year}{2020}), \eprint{2001.00922}.

\bibitem[{\citenamefont{Abeysekara et~al.}(2017)\citenamefont{Abeysekara,
  Albert, Alfaro, Alvarez, {\'{A}}lvarez, Arceo, Arteaga-Vel{\'{a}}zquez,
  Solares, Barber, Bautista-Elivar et~al.}}]{hawc}
\bibinfo{author}{\bibfnamefont{A.~U.} \bibnamefont{Abeysekara}},
  \bibinfo{author}{\bibfnamefont{A.}~\bibnamefont{Albert}},
  \bibinfo{author}{\bibfnamefont{R.}~\bibnamefont{Alfaro}},
  \bibinfo{author}{\bibfnamefont{C.}~\bibnamefont{Alvarez}},
  \bibinfo{author}{\bibfnamefont{J.~D.} \bibnamefont{{\'{A}}lvarez}},
  \bibinfo{author}{\bibfnamefont{R.}~\bibnamefont{Arceo}},
  \bibinfo{author}{\bibfnamefont{J.~C.} \bibnamefont{Arteaga-Vel{\'{a}}zquez}},
  \bibinfo{author}{\bibfnamefont{H.~A.~A.} \bibnamefont{Solares}},
  \bibinfo{author}{\bibfnamefont{A.~S.} \bibnamefont{Barber}},
  \bibinfo{author}{\bibfnamefont{N.}~\bibnamefont{Bautista-Elivar}},
  \bibnamefont{et~al.}, \bibinfo{journal}{The Astrophysical Journal}
  \textbf{\bibinfo{volume}{843}}, \bibinfo{pages}{39} (\bibinfo{year}{2017}),
  \urlprefix\url{https://doi.org/10.3847%2F1538-4357%2Faa7555}.

\bibitem[{\citenamefont{{Wu} et~al.}(2019)\citenamefont{{Wu}, {Chen}, {He},
  {Lin}, and {Nan}}}]{lhaaso_electrons}
\bibinfo{author}{\bibfnamefont{S.}~\bibnamefont{{Wu}}},
  \bibinfo{author}{\bibfnamefont{S.}~\bibnamefont{{Chen}}},
  \bibinfo{author}{\bibfnamefont{H.}~\bibnamefont{{He}}},
  \bibinfo{author}{\bibfnamefont{S.}~\bibnamefont{{Lin}}}, \bibnamefont{and}
  \bibinfo{author}{\bibfnamefont{Y.}~\bibnamefont{{Nan}}}, in
  \emph{\bibinfo{booktitle}{36th International Cosmic Ray Conference
  (ICRC2019)}} (\bibinfo{year}{2019}), vol.~\bibinfo{volume}{36} of
  \emph{\bibinfo{series}{International Cosmic Ray Conference}}, p.
  \bibinfo{pages}{471}.

\bibitem[{\citenamefont{{Abeysekara} et~al.}(2018)\citenamefont{{Abeysekara},
  {Albert}, {Alfaro}, {Alvarez}, {Arceo}, {Arteaga-Vel{\'a}zquez}, {Avila
  Rojas}, {Ayala Solares}, {Becerril}, {Belmont-Moreno} et~al.}}]{hawc_dm}
\bibinfo{author}{\bibfnamefont{A.~U.} \bibnamefont{{Abeysekara}}},
  \bibinfo{author}{\bibfnamefont{A.}~\bibnamefont{{Albert}}},
  \bibinfo{author}{\bibfnamefont{R.}~\bibnamefont{{Alfaro}}},
  \bibinfo{author}{\bibfnamefont{C.}~\bibnamefont{{Alvarez}}},
  \bibinfo{author}{\bibfnamefont{R.}~\bibnamefont{{Arceo}}},
  \bibinfo{author}{\bibfnamefont{J.~C.} \bibnamefont{{Arteaga-Vel{\'a}zquez}}},
  \bibinfo{author}{\bibfnamefont{D.}~\bibnamefont{{Avila Rojas}}},
  \bibinfo{author}{\bibfnamefont{H.~A.} \bibnamefont{{Ayala Solares}}},
  \bibinfo{author}{\bibfnamefont{A.}~\bibnamefont{{Becerril}}},
  \bibinfo{author}{\bibfnamefont{E.}~\bibnamefont{{Belmont-Moreno}}},
  \bibnamefont{et~al.}, \bibinfo{journal}{\jcap}
  \textbf{\bibinfo{volume}{2018}}, \bibinfo{eid}{049} (\bibinfo{year}{2018}),
  \eprint{1710.10288}.

\bibitem[{\citenamefont{{Abeysekara} et~al.}(2017)\citenamefont{{Abeysekara},
  {Albert}, {Alfaro}, {Alvarez}, {{\'A}lvarez}, {Arceo},
  {Arteaga-Vel{\'a}zquez}, {Ayala Solares}, {Barber}, {Bautista-Elivar}
  et~al.}}]{hawc_bubble}
\bibinfo{author}{\bibfnamefont{A.~U.} \bibnamefont{{Abeysekara}}},
  \bibinfo{author}{\bibfnamefont{A.}~\bibnamefont{{Albert}}},
  \bibinfo{author}{\bibfnamefont{R.}~\bibnamefont{{Alfaro}}},
  \bibinfo{author}{\bibfnamefont{C.}~\bibnamefont{{Alvarez}}},
  \bibinfo{author}{\bibfnamefont{J.~D.} \bibnamefont{{{\'A}lvarez}}},
  \bibinfo{author}{\bibfnamefont{R.}~\bibnamefont{{Arceo}}},
  \bibinfo{author}{\bibfnamefont{J.~C.} \bibnamefont{{Arteaga-Vel{\'a}zquez}}},
  \bibinfo{author}{\bibfnamefont{H.~A.} \bibnamefont{{Ayala Solares}}},
  \bibinfo{author}{\bibfnamefont{A.~S.} \bibnamefont{{Barber}}},
  \bibinfo{author}{\bibfnamefont{N.}~\bibnamefont{{Bautista-Elivar}}},
  \bibnamefont{et~al.}, \bibinfo{journal}{\apj} \textbf{\bibinfo{volume}{842}},
  \bibinfo{eid}{85} (\bibinfo{year}{2017}), \eprint{1703.01344}.

\bibitem[{\citenamefont{{Williams}}(2019)}]{icrc2019}
\bibinfo{author}{\bibfnamefont{D.}~\bibnamefont{{Williams}}}, in
  \emph{\bibinfo{booktitle}{36th International Cosmic Ray Conference
  (ICRC2019)}} (\bibinfo{year}{2019}), vol.~\bibinfo{volume}{36} of
  \emph{\bibinfo{series}{International Cosmic Ray Conference}},
  p.~\bibinfo{pages}{16}, \eprint{1909.05173}.

\bibitem[{\citenamefont{{Harding}}(2019)}]{hawc_diffuse}
\bibinfo{author}{\bibfnamefont{J.~P.} \bibnamefont{{Harding}}}, in
  \emph{\bibinfo{booktitle}{36th International Cosmic Ray Conference
  (ICRC2019)}} (\bibinfo{year}{2019}), vol.~\bibinfo{volume}{36} of
  \emph{\bibinfo{series}{International Cosmic Ray Conference}}, p.
  \bibinfo{pages}{691}, \eprint{1908.11485}.

\bibitem[{\citenamefont{Abeysekara et~al.}(2018)\citenamefont{Abeysekara,
  Alfaro, Alvarez, {\'{A}}lvarez, Arceo, Arteaga-Vel{\'{a}}zquez, Rojas,
  Solares, Becerril, Belmont-Moreno et~al.}}]{cr_anisotropy}
\bibinfo{author}{\bibfnamefont{A.~U.} \bibnamefont{Abeysekara}},
  \bibinfo{author}{\bibfnamefont{R.}~\bibnamefont{Alfaro}},
  \bibinfo{author}{\bibfnamefont{C.}~\bibnamefont{Alvarez}},
  \bibinfo{author}{\bibfnamefont{J.~D.} \bibnamefont{{\'{A}}lvarez}},
  \bibinfo{author}{\bibfnamefont{R.}~\bibnamefont{Arceo}},
  \bibinfo{author}{\bibfnamefont{J.~C.} \bibnamefont{Arteaga-Vel{\'{a}}zquez}},
  \bibinfo{author}{\bibfnamefont{D.~A.} \bibnamefont{Rojas}},
  \bibinfo{author}{\bibfnamefont{H.~A.~A.} \bibnamefont{Solares}},
  \bibinfo{author}{\bibfnamefont{A.}~\bibnamefont{Becerril}},
  \bibinfo{author}{\bibfnamefont{E.}~\bibnamefont{Belmont-Moreno}},
  \bibnamefont{et~al.}, \bibinfo{journal}{The Astrophysical Journal}
  \textbf{\bibinfo{volume}{865}}, \bibinfo{pages}{57} (\bibinfo{year}{2018}),
  \urlprefix\url{https://doi.org/10.3847%2F1538-4357%2Faad90c}.

\bibitem[{\citenamefont{Kachelriess and Semikoz}(2019)}]{review_CR}
\bibinfo{author}{\bibfnamefont{M.}~\bibnamefont{Kachelriess}} \bibnamefont{and}
  \bibinfo{author}{\bibfnamefont{D.~V.} \bibnamefont{Semikoz}},
  \bibinfo{journal}{Prog. Part. Nucl. Phys.} \textbf{\bibinfo{volume}{109}},
  \bibinfo{pages}{103710} (\bibinfo{year}{2019}), \eprint{1904.08160}.

\bibitem[{\citenamefont{{Syrovatskii}}(1971)}]{1971CoASP...3..155S}
\bibinfo{author}{\bibfnamefont{S.~I.} \bibnamefont{{Syrovatskii}}},
  \bibinfo{journal}{Comments on Astrophysics and Space Physics}
  \textbf{\bibinfo{volume}{3}}, \bibinfo{pages}{155} (\bibinfo{year}{1971}).

\bibitem[{\citenamefont{{Ptuskin} et~al.}(1993)\citenamefont{{Ptuskin},
  {Rogovaya}, {Zirakashvili}, {Chuvilgin}, {Khristiansen}, {Klepach}, and
  {Kulikov}}}]{1993A&A...268..726P}
\bibinfo{author}{\bibfnamefont{V.~S.} \bibnamefont{{Ptuskin}}},
  \bibinfo{author}{\bibfnamefont{S.~I.} \bibnamefont{{Rogovaya}}},
  \bibinfo{author}{\bibfnamefont{V.~N.} \bibnamefont{{Zirakashvili}}},
  \bibinfo{author}{\bibfnamefont{L.~G.} \bibnamefont{{Chuvilgin}}},
  \bibinfo{author}{\bibfnamefont{G.~B.} \bibnamefont{{Khristiansen}}},
  \bibinfo{author}{\bibfnamefont{E.~G.} \bibnamefont{{Klepach}}},
  \bibnamefont{and} \bibinfo{author}{\bibfnamefont{G.~V.}
  \bibnamefont{{Kulikov}}}, \bibinfo{journal}{\aap}
  \textbf{\bibinfo{volume}{268}}, \bibinfo{pages}{726} (\bibinfo{year}{1993}).

\bibitem[{\citenamefont{Candia et~al.}(2002)\citenamefont{Candia, Roulet, and
  Epele}}]{Candia:2002we}
\bibinfo{author}{\bibfnamefont{J.}~\bibnamefont{Candia}},
  \bibinfo{author}{\bibfnamefont{E.}~\bibnamefont{Roulet}}, \bibnamefont{and}
  \bibinfo{author}{\bibfnamefont{L.~N.} \bibnamefont{Epele}},
  \bibinfo{journal}{JHEP} \textbf{\bibinfo{volume}{12}}, \bibinfo{pages}{033}
  (\bibinfo{year}{2002}), \eprint{astro-ph/0206336}.

\bibitem[{\citenamefont{Giacinti et~al.}(2014)\citenamefont{Giacinti,
  Kachelrieß, and Semikoz}}]{Giacinti:2014xya}
\bibinfo{author}{\bibfnamefont{G.}~\bibnamefont{Giacinti}},
  \bibinfo{author}{\bibfnamefont{M.}~\bibnamefont{Kachelrieß}},
  \bibnamefont{and} \bibinfo{author}{\bibfnamefont{D.~V.}
  \bibnamefont{Semikoz}}, \bibinfo{journal}{Phys. Rev.}
  \textbf{\bibinfo{volume}{D90}}, \bibinfo{pages}{041302}
  (\bibinfo{year}{2014}), \eprint{1403.3380}.

\bibitem[{\citenamefont{Giacinti et~al.}(2015)\citenamefont{Giacinti,
  Kachelrieß, and Semikoz}}]{Giacinti:2015hva}
\bibinfo{author}{\bibfnamefont{G.}~\bibnamefont{Giacinti}},
  \bibinfo{author}{\bibfnamefont{M.}~\bibnamefont{Kachelrieß}},
  \bibnamefont{and} \bibinfo{author}{\bibfnamefont{D.~V.}
  \bibnamefont{Semikoz}}, \bibinfo{journal}{Phys. Rev.}
  \textbf{\bibinfo{volume}{D91}}, \bibinfo{pages}{083009}
  (\bibinfo{year}{2015}), \eprint{1502.01608}.

\bibitem[{\citenamefont{Giacinti et~al.}(2018)\citenamefont{Giacinti,
  Kachelriess, and Semikoz}}]{Giacinti:2017dgt}
\bibinfo{author}{\bibfnamefont{G.}~\bibnamefont{Giacinti}},
  \bibinfo{author}{\bibfnamefont{M.}~\bibnamefont{Kachelriess}},
  \bibnamefont{and} \bibinfo{author}{\bibfnamefont{D.~V.}
  \bibnamefont{Semikoz}}, \bibinfo{journal}{JCAP}
  \textbf{\bibinfo{volume}{1807}}, \bibinfo{pages}{051} (\bibinfo{year}{2018}),
  \eprint{1710.08205}.

\bibitem[{\citenamefont{{Jansson} and {Farrar}}(2012)}]{jansson}
\bibinfo{author}{\bibfnamefont{R.}~\bibnamefont{{Jansson}}} \bibnamefont{and}
  \bibinfo{author}{\bibfnamefont{G.~R.} \bibnamefont{{Farrar}}},
  \bibinfo{journal}{\apj} \textbf{\bibinfo{volume}{757}}, \bibinfo{eid}{14}
  (\bibinfo{year}{2012}), \eprint{1204.3662}.

\bibitem[{\citenamefont{Stanev et~al.}(1993)\citenamefont{Stanev, Biermann, and
  Gaisser}}]{Stanev:1993tx}
\bibinfo{author}{\bibfnamefont{T.}~\bibnamefont{Stanev}},
  \bibinfo{author}{\bibfnamefont{P.~L.} \bibnamefont{Biermann}},
  \bibnamefont{and} \bibinfo{author}{\bibfnamefont{T.~K.}
  \bibnamefont{Gaisser}}, \bibinfo{journal}{Astron. Astrophys.}
  \textbf{\bibinfo{volume}{274}}, \bibinfo{pages}{902} (\bibinfo{year}{1993}),
  \eprint{astro-ph/9303006}.

\bibitem[{\citenamefont{Kobayakawa et~al.}(2002)\citenamefont{Kobayakawa, Sato,
  and Samura}}]{Kobayakawa:2000nq}
\bibinfo{author}{\bibfnamefont{K.}~\bibnamefont{Kobayakawa}},
  \bibinfo{author}{\bibfnamefont{Y.}~\bibnamefont{Sato}}, \bibnamefont{and}
  \bibinfo{author}{\bibfnamefont{T.}~\bibnamefont{Samura}},
  \bibinfo{journal}{Phys. Rev.} \textbf{\bibinfo{volume}{D66}},
  \bibinfo{pages}{083004} (\bibinfo{year}{2002}), \eprint{astro-ph/0008209}.

\bibitem[{\citenamefont{Hillas}(2005)}]{Hillas_2005}
\bibinfo{author}{\bibfnamefont{A.~M.} \bibnamefont{Hillas}},
  \bibinfo{journal}{Journal of Physics G: Nuclear and Particle Physics}
  \textbf{\bibinfo{volume}{31}}, \bibinfo{pages}{R95} (\bibinfo{year}{2005}),
  \urlprefix\url{https://doi.org/10.1088%2F0954-3899%2F31%2F5%2Fr02}.

\bibitem[{\citenamefont{Zatsepin and Sokolskaya}(2006)}]{Zatsepin:2006ci}
\bibinfo{author}{\bibfnamefont{V.~I.} \bibnamefont{Zatsepin}} \bibnamefont{and}
  \bibinfo{author}{\bibfnamefont{N.~V.} \bibnamefont{Sokolskaya}},
  \bibinfo{journal}{Astron. Astrophys.} \textbf{\bibinfo{volume}{458}},
  \bibinfo{pages}{1} (\bibinfo{year}{2006}), \eprint{astro-ph/0601475}.

\bibitem[{\citenamefont{Erlykin and Wolfendale}(1997)}]{Erlykin:1997bs}
\bibinfo{author}{\bibfnamefont{A.~D.} \bibnamefont{Erlykin}} \bibnamefont{and}
  \bibinfo{author}{\bibfnamefont{A.~W.} \bibnamefont{Wolfendale}},
  \bibinfo{journal}{J. Phys.} \textbf{\bibinfo{volume}{G23}},
  \bibinfo{pages}{979} (\bibinfo{year}{1997}).

\bibitem[{\citenamefont{Erlykin and Wolfendale}(2001)}]{Erlykin:2000jm}
\bibinfo{author}{\bibfnamefont{A.~D.} \bibnamefont{Erlykin}} \bibnamefont{and}
  \bibinfo{author}{\bibfnamefont{A.~W.} \bibnamefont{Wolfendale}},
  \bibinfo{journal}{Adv. Space Res.} \textbf{\bibinfo{volume}{27}},
  \bibinfo{pages}{803} (\bibinfo{year}{2001}), \eprint{astro-ph/0011057}.

\bibitem[{\citenamefont{Erlykin and Wolfendale}(2015)}]{Erlykin2015}
\bibinfo{author}{\bibfnamefont{A.~D.} \bibnamefont{Erlykin}} \bibnamefont{and}
  \bibinfo{author}{\bibfnamefont{A.~W.} \bibnamefont{Wolfendale}},
  \bibinfo{journal}{Bulletin of the Russian Academy of Sciences: Physics}
  \textbf{\bibinfo{volume}{79}}, \bibinfo{pages}{308} (\bibinfo{year}{2015}),
  ISSN \bibinfo{issn}{1934-9432},
  \urlprefix\url{https://doi.org/10.3103/S1062873815030181}.

\bibitem[{\citenamefont{{Boyarsky} et~al.}(2006)\citenamefont{{Boyarsky},
  {Neronov}, {Ruchayskiy}, {Shaposhnikov}, and {Tkachev}}}]{boyarsky06}
\bibinfo{author}{\bibfnamefont{A.}~\bibnamefont{{Boyarsky}}},
  \bibinfo{author}{\bibfnamefont{A.}~\bibnamefont{{Neronov}}},
  \bibinfo{author}{\bibfnamefont{O.}~\bibnamefont{{Ruchayskiy}}},
  \bibinfo{author}{\bibfnamefont{M.}~\bibnamefont{{Shaposhnikov}}},
  \bibnamefont{and}
  \bibinfo{author}{\bibfnamefont{I.}~\bibnamefont{{Tkachev}}},
  \bibinfo{journal}{\prl} \textbf{\bibinfo{volume}{97}}, \bibinfo{eid}{261302}
  (\bibinfo{year}{2006}), \eprint{astro-ph/0603660}.

\bibitem[{\citenamefont{{Boyarsky} et~al.}(2007)\citenamefont{{Boyarsky}, {den
  Herder}, {Neronov}, and {Ruchayskiy}}}]{boyarsky07}
\bibinfo{author}{\bibfnamefont{A.}~\bibnamefont{{Boyarsky}}},
  \bibinfo{author}{\bibfnamefont{J.~W.} \bibnamefont{{den Herder}}},
  \bibinfo{author}{\bibfnamefont{A.}~\bibnamefont{{Neronov}}},
  \bibnamefont{and}
  \bibinfo{author}{\bibfnamefont{O.}~\bibnamefont{{Ruchayskiy}}},
  \bibinfo{journal}{Astroparticle Physics} \textbf{\bibinfo{volume}{28}},
  \bibinfo{pages}{303} (\bibinfo{year}{2007}), \eprint{astro-ph/0612219}.

\bibitem[{\citenamefont{{Boyarsky} et~al.}(2008)\citenamefont{{Boyarsky},
  {Malyshev}, {Neronov}, and {Ruchayskiy}}}]{boyarsky08}
\bibinfo{author}{\bibfnamefont{A.}~\bibnamefont{{Boyarsky}}},
  \bibinfo{author}{\bibfnamefont{D.}~\bibnamefont{{Malyshev}}},
  \bibinfo{author}{\bibfnamefont{A.}~\bibnamefont{{Neronov}}},
  \bibnamefont{and}
  \bibinfo{author}{\bibfnamefont{O.}~\bibnamefont{{Ruchayskiy}}},
  \bibinfo{journal}{\mnras} \textbf{\bibinfo{volume}{387}},
  \bibinfo{pages}{1345} (\bibinfo{year}{2008}), \eprint{0710.4922}.

\bibitem[{\citenamefont{{Klypin} et~al.}(2002)\citenamefont{{Klypin}, {Zhao},
  and {Somerville}}}]{klypin}
\bibinfo{author}{\bibfnamefont{A.}~\bibnamefont{{Klypin}}},
  \bibinfo{author}{\bibfnamefont{H.}~\bibnamefont{{Zhao}}}, \bibnamefont{and}
  \bibinfo{author}{\bibfnamefont{R.~S.} \bibnamefont{{Somerville}}},
  \bibinfo{journal}{\apj} \textbf{\bibinfo{volume}{573}}, \bibinfo{pages}{597}
  (\bibinfo{year}{2002}), \eprint{astro-ph/0110390}.

\bibitem[{\citenamefont{{Kachelrie{\ss}}
  et~al.}(2018)\citenamefont{{Kachelrie{\ss}}, {Kalashev}, and
  {Kuznetsov}}}]{kalashev}
\bibinfo{author}{\bibfnamefont{M.}~\bibnamefont{{Kachelrie{\ss}}}},
  \bibinfo{author}{\bibfnamefont{O.~E.} \bibnamefont{{Kalashev}}},
  \bibnamefont{and} \bibinfo{author}{\bibfnamefont{M.~Y.}
  \bibnamefont{{Kuznetsov}}}, \bibinfo{journal}{\prd}
  \textbf{\bibinfo{volume}{98}}, \bibinfo{eid}{083016} (\bibinfo{year}{2018}),
  \eprint{1805.04500}.

\end{thebibliography}
\end{document}